\documentclass[11pt,onecolumn,a4paper]{article}

\usepackage[top=0.75in,bottom=1in,left=0.9in,right=0.9in]{geometry}
\usepackage{amsmath}
\usepackage{amssymb}
\usepackage{bm}
\usepackage{booktabs}
\usepackage{titling}
\usepackage{xcolor}
\usepackage{graphicx}
\newcommand{\figscale}{1.2}
\graphicspath{{figures/}}
\usepackage[style=numeric,giveninits=true,uniquename=init,minnames=1,maxnames=100,doi=false,url=false,isbn=false,natbib=true]{biblatex}
\DeclareFieldFormat{title}{#1}
\AtBeginBibliography{\small}

\renewcommand{\vec}[1]{\boldsymbol{#1}}

\newcommand{\pderiv}[2]{\frac{\partial #1}{\partial #2}}
\newcommand{\Wi}{\mathrm{Wi}}
\renewcommand{\Re}{\mathrm{Re}}

\usepackage{authblk}

\author[1]{Calum S.\ Skene}
\author[2]{Jeffrey S. Oishi}
\author[3,4]{Keaton J.\ Burns}
\author[5]{Moritz Linkmann}
\author[5]{Geoffrey M. Vasil}
\author[1]{Alexander Morozov}
\affil[1]{School of Physics and Astronomy, The University of Edinburgh, James Clerk Maxwell Building, Peter Guthrie Tait Road, Edinburgh, EH9 3FD, United Kingdom}
\affil[2]{Program in Integrated Applied Mathematics and Department of Mechanical Engineering, University of New Hampshire, Durham NH 03824, USA}
\affil[3]{Department of Mathematics, Massachusetts Institute of Technology, Cambridge MA 02139, USA}
\affil[4]{Center for Computational Astrophysics, Flatiron Institute, New York NY 10010, USA}
\affil[5]{School of Mathematics and the Maxwell Institute for Mathematical Sciences, The University of Edinburgh, Edinburgh EH9 3FD, United Kingdom}
\begin{document}

% Title
\title{\Large\bfseries Purely elastic centre-mode instability in channel and pipe flows of a simplified Phan-Thien--Tanner fluid}
\date{\vspace{-2em}}

\maketitle
\begin{abstract}
\normalsize
Recent work identified spatially localised midplane stress structures as the building blocks of chaotic behaviour in pressure-driven parallel shear flows of viscoelastic fluids. Such coherent structures are thought to originate from a linear centre-mode instability of the laminar state. In the absence of inertia, however, the Oldroyd-B and FENE-P constitutive models predict that this instability is confined to ultra-dilute polymer solutions and very large Weissenberg numbers in channel flow, while no purely elastic instability has been reported in pipe flow. Here, we study numerically the linear stability of channel and pipe flows of a simplified Phan-Thien--Tanner (sPTT) fluid and find a purely elastic centre-mode instability in both geometries across a wide range of polymer concentrations and at much lower Weissenberg numbers. Since the sPTT and FENE-P models have identical steady rheology and laminar profiles, their contrasting predictions indicate that the existence of the centre-mode instability is governed by the transient rheological response of the fluid. Our results suggest that the purely elastic centre-mode instability should be observable under realistic flow conditions. They also offer a means of discriminating between constitutive models.
\end{abstract}

% Main body
\section{Introduction}\label{sec:intro}

Recent years have seen significant progress in understanding elastic turbulence in parallel shear flows. Through large-scale direct numerical simulations it has been shown that elastic turbulence is organised around the midplane/centreline of such geometries \citep{Beneitez2024,Lellep2024,Rota2024,Lin2025}, unlike Newtonian turbulence, which is dominated by near-wall structures \citep{Smits2011}. The associated three-dimensional coherent states comprise pairs of vortices, oriented predominantly perpendicular to the direction of the flow, that extrude midplane regions of very high polymer stretch \citep{Lellep2024,Morozov2025preprint}. Such localised structures are organised by two-dimensional travelling-wave solutions: the \emph{arrowheads}, in the elasto-inertial regime, and \emph{narwhals}, in the purely elastic regime, found in channel, pipe, and Kolmogorov flows \citep{Berti2008,Berti2010,Page2020,Morozov2022,Beneitez2024,Nichols2025,Lewy2025,Burns2026preprint}, that connect to the three-dimensional structures through a spanwise linear instability \citep{Lellep2023}. Where available, systematic continuation of the two-dimensional solutions indicates that they bifurcate from a linear instability \citep{Page2020,Buza2022,Buza2022a,Zhu2026}, and it is therefore of prime importance to understand the conditions under which such an instability can take place. 

The linear instability in question, the so-called centre mode instability, is a novel phenomenon only recently discovered in parallel shear flows of viscoelastic fluids \citep{Datta2022,Sanchez2022}. Until then, the only known purely elastic linear instability was associated with the hoop stress mechanism \citep{Shaqfeh1996,McKinley1996} that requires curved streamlines in the base flow. While the hoop-stress mode becomes stable in the limit of parallel shear flows \citep{Wilson1999}, the centre-mode instability is rather insensitive to the geometry curvature \citep{Tej2026} and persists in that limit. While a matched-asymptotics analysis identifies the centre-mode as a critical-layer type instability \citep{Kerswell2024}, there is currently no mechanistic understanding of its structure, of the kind recently proposed for the narwhal and arrowhead solutions \citep{Morozov2022,Morozov2025preprint,Goffin2026}.

The region of existence of the centre-mode instability has been mapped out in great detail in pipe and channel flows of the Oldroyd-B and FENE-P fluids. Its boundaries are expressed in terms of the Reynolds number, $\Re=\rho U H/(\eta_s+\eta_p)$, which measures the importance of the fluid inertia, the Weissenberg number, $\Wi = \lambda U/H$, which measures the fluid elasticity, and the viscosity ratio, $\beta = \eta_s/(\eta_s+\eta_p)$, with $1-\beta$ serving as a proxy for the polymer concentration \citep{Bird1987}. Here, $U$ is the laminar centreline velocity of the fluid, $H$ is the half-width or the radius, in channel and pipe flows, respectively, $\lambda$ is the polymer relaxation time, $\eta_s$ and $\eta_p$ are the solvent and polymer contributions to the total viscosity, and $\rho$ is the density. For the Oldroyd-B model, which exhibits rate-independent material properties, the instability was first reported at $\Re\sim O(10^2$--$10^3)$ \citep{Garg2018,Chaudhary2021,Khalid2021a}, and its purely elastic counterpart occupies a remarkably small part of the parameter space. In channel flow, the instability persists as $\Re\to 0$ only for ultra-dilute solutions, $\beta>0.9905$, and for Weissenberg numbers $\Wi\sim O(10^3)$: the lowest critical value, $\Wi_c\approx 974$, is reached for $\beta\approx0.994$, with $\Wi_c$ rising sharply as $\beta\to 0.9905$ and growing as $(1-\beta)^{-1}$ as $\beta\to 1$ \citep{Khalid2021,Kerswell2024}. At lower $\beta$ a finite amount of inertia is required to destabilise the flow, and for $\beta\lesssim 0.5$ the instability is absent altogether \citep{Khalid2021a}. In pipe flow, the instability has only been found for $\Re\gtrsim 63$ \citep{Chaudhary2021}, and no purely elastic centre mode has been reported. 

For other models, to our knowledge, the available results for parallel shear flows are limited to channel flow of FENE-type fluids. There, the effect of finite extensibility was found to be only mildly destabilising in the purely elastic regime: for a maximum extension $L=O(100)$, the critical Weissenberg number drops to $\Wi_c\approx 165$, and the instability extends down to $\beta\approx 0.97$ \citep{Buza2022,Khalid2025}. This effect is non-monotonic in $L$, and is reversed at higher polymer concentrations, where finite extensibility is stabilising and no instability is found at small $\Re$ \citep{Buza2022}. It is also absent in the FENE-CR model, where finite extensibility suppresses the elastic instability \citep{Khalid2025}. 

The current theoretical understanding is therefore that, in purely elastic channel flows, the linear centre-mode instability is confined to very dilute solutions, with $\beta\gtrsim 0.97$ and $\Wi\gtrsim O(10^2)$, while in pipe flows, no such instability has been reported for $\Re\ll 1$.

Here, we present the surprising result that the centre-mode instability can be found for a broad range of parameters in purely elastic pipe and channel flows provided one uses the  simplified Phan-Thien-Tanner (sPTT) constitutive equation \citep{PhanThien1977} as a polymer model. This finding is unexpected as the sPTT model’s rheological predictions in steady and homogeneous flows are identical to those of the FENE-P model \citep{Oliveira1999,John2024}. Below we map out the region of existence of the centre-mode instability in channel and pipe flows of an sPTT fluid and discuss the implications of these results for the theoretical understanding of elastic instabilities and turbulence.

\vspace*{-0.4cm}
\section{Mathematical and numerical formulation}
We study incompressible, pressure-driven flows of a model viscoelastic fluid through a straight channel formed by two infinite parallel plates, or a pipe of circular cross-section. To describe these geometries, we introduce Cartesian, $(x, \eta, z)$, and cylindrical polar, $(\eta, \theta, x)$, coordinate systems, with $x$ being the streamwise coordinate in both cases. 
The spanwise and azimuthal coordinates are $z$ and $\theta$, respectively, while $\eta$ denotes the wall normal direction in the channel, and the radial coordinate in a pipe.
The dynamics of the fluid is governed by the simplified Phan-Thien-Tanner model \citep{PhanThien1977} given by
\begin{gather}
\label{nseq}
\Re\left(\pderiv{\bm{u}}{t} + \bm{u} \cdot\nabla\bm{u}\right) = -\nabla p + \beta\nabla^2\bm{u} + \frac{1-\beta}{\Wi}\nabla\cdot\bm{c}+ \Delta p\bm{e}_x, \\
\label{incompeq}
\nabla\cdot\bm{u} = 0 ,\\
\label{sptteq}
\pderiv{\bm{c}}{t} + \vec{u} \cdot\nabla\bm{c} - (\nabla\bm{u})^T\cdot\bm{c} - \bm{c} \cdot \nabla\bm{u} = 
- \frac{\bm{c}-\vec{I}}{\Wi}\left[1+\epsilon\textrm{Tr}(\bm{c}-\bm{I}) \right].
\end{gather}
Here, $\bm{u}$ is the fluid velocity, $\bm{c}$ is the polymer conformation tensor, and $p$ is the pressure. The flow is driven by an externally applied pressure gradient, $\Delta p=2$, for a channel, and $\Delta p=4$, for pipe, $\bm{e}_x$ is a unit vector in the streamwise direction, $\bm{I}$ is the identity tensor, while $\textrm{Tr}(\cdots)$ and $(\cdots)^T$ denote the trace and the transpose, respectively. These equations are rendered dimensionless by rescaling all lengths with $H$, time with $H/U$, and the velocity with $U$ - the maximum velocity of a Newtonian fluid with a matching viscosity and driven by the same pressure gradient. The degree of shear thinning is controlled by $\epsilon$. Finally, the boundary conditions are given by $\bm{u}(\eta=\pm1)=0$, in channel, and $\bm{u}(\eta=1)=0$ with $\bm{u}(\eta=0)$ being finite, in pipe flows.

Eqs.\eqref{nseq}-\eqref{sptteq} admit an analytic, unidirectional and steady solution, $\{\bm{c},\bm{u}\}({\bm{r},t})=\{\bm{C},U \bm{e}_x\}(\eta)$, where
\begin{align}
\label{Cxeta_lam}
& C_{x\eta}(\eta) = -\sqrt{\frac{2}{3\epsilon\beta}} \sinh \left[\frac{1}{3}\mathrm{arcsinh}\left(3\Wi\,\eta\sqrt{6\epsilon\beta}\right)  \right],
\end{align}
\begin{align}
\label{Cxx_lam}
& C_{xx}(\eta) = 1+ 2C_{x\eta}(\eta)^2,  \\
\label{U_lam}
& U(\eta) = 1-\eta^2 + \epsilon\frac{1-\beta}{4\Wi^2}[G(C_{x\eta}(1)) - G(C_{x\eta}(\eta))],
\end{align}
and $G(s) = s^4(1+4\epsilon\beta s^2)$. Additionally, $C_{zz}=1$ (channel) and $C_{\theta\theta}=1$ (pipe), $C_{\eta\eta} = 1$ in both geometries, while all other components of the conformation tensor are zero. We note that this solution is equivalent to that obtained by~\citet{Cruz2005}, who gave the solution in terms of radicals.

To analyse linear stability of the laminar state given above, we consider time evolution of infinitesimal perturbations in the form 
\begin{align}
\{\bm{c},\bm{u},p\}({\bm{r},t})=\{\bm{C},U \bm{e}_x, P\}(\eta) + e^{(\sigma_r + i \sigma_i) t + i k_x x} A \{\bm{\delta c},\bm{\delta u},\delta p\}(\eta) + c.c.,
\end{align}
where $A=e^{i k_z z}$, for channel flow, and $A=e^{i m \theta}$, for pipe flow. Here, $k_x$ and $k_z$ denote the streamwise and spanwise wavenumbers, respectively, $m$ is an (integer) azimuthal wavenumber, $\sigma_r$ and $\sigma_i$ are the real and imaginary part of the temporal eigenvalue, respectively, and $c.c.$ denotes the complex conjugate. Linearising Eqs.\eqref{nseq}-\eqref{sptteq} around the laminar profile yields a generalised eigenvalue problem for $\sigma_r + i \sigma_i$, which we solve numerically. An instability is identified when $\sigma_r>0$. While Squire's theorem guarantees that the most unstable modes are two-dimensional in channel flows of Newtonian and Oldroyd-B fluids \citep{Bistagnino_2007}, it was shown not to hold for the sPTT model \citep{Furlan2022}. Indeed, below we demonstrate that the most unstable perturbations in channel flow correspond to $k_z\neq 0$. In pipe flow, on the other hand, we find that only perturbations with $m=0$ are unstable, and below we restrict our results to that case.

The generalised eigenvalue problem for $\sigma$ is solved numerically using the open-source spectral framework Dedalus~\citep{Burns2020}. The discretisation uses Chebyshev polynomials in the channel and Zernike polynomials in the pipe~\citep{Vasil2016}. Unstable modes were initially identified by performing a coarse scan over $\mathrm{Wi}$ and wavenumber. The instabilities were then tracked using pseudo-arclength continuation in ($\mathrm{Wi}$, $\beta$), or ($\mathrm{Wi}$, $\mathrm{Re}$), combined with optimisation using Brent's method in 2D and Newton's method in 3D to maximise growth over wavenumber. The resolution was chosen for each parameter set such that the eigenvector's spectral coefficients decayed sufficiently and the critical eigenvalue remained fixed when the resolution was increased by a factor of $3/2$.

\section{Results}\label{sec:results}

\begin{figure}[!t]
\begin{center}
\includegraphics[scale=\figscale]{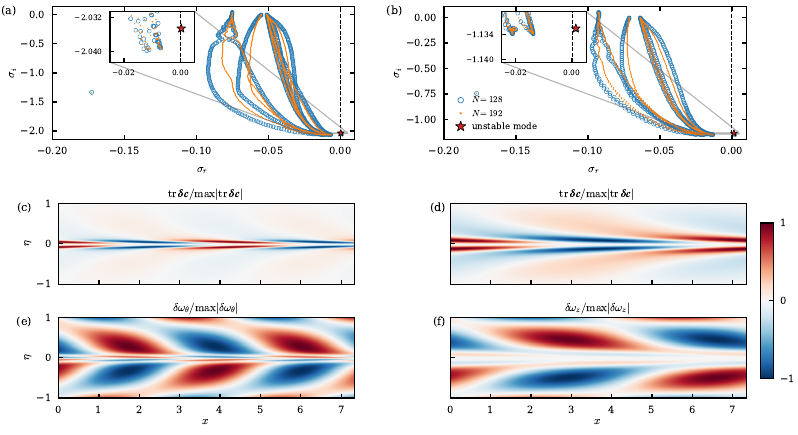}
\end{center}
\caption{Unstable centre mode of purely elastic flow of an sPTT fluid for $\Re=0$, $\beta=0.65$, and $\epsilon=10^{-3}$.  Eigenvalue spectra for (a) pipe flow, $m=0$, $\Wi=138.4$, $k_x=1.470$, and (b) channel flow, $\Wi=72.6$, $k_x=0.856$, $k_z=0.948$; the unstable mode is marked by a star, and the insets show the region near $\sigma_r=0$. Symbols of different colour denote eigenvalues calculated at two spatial resolutions.
(c--f) Corresponding eigenfunctions: perturbation of the polymer stretch, $\textrm{Tr}\,\bm{\delta c}$ (c--d), and perturbation vorticity (e--f) in pipe flow (c,e), and in channel flow in the streamwise--wall-normal plane $z=0$ (d,f). Each field is normalised by its maximum absolute value.}
\label{fig1}
\end{figure}

\subsection{The unstable mode}\label{sec:mode}

We begin by establishing that the laminar flow of an sPTT fluid is linearly unstable in the absence of inertia, $\Re=0$, in both geometries. Representative eigenvalue spectra are shown in Fig.~\ref{fig1}(a,b). Besides the continuous spectra, each spectrum contains a single unstable discrete eigenvalue, whose phase speed, $c=-\sigma_i/k_x$, is within a fraction of a per cent of the maximum velocity of the laminar flow. 
The unstable mode therefore sits next to the centreline, in the region of nearly vanishing base shear, which identifies it as a centre mode. In pipe flow, the unstable mode is axisymmetric, while in channel flow the most unstable perturbation is three-dimensional, with $k_z\approx k_x$ for the parameters of Fig.~\ref{fig1}(b).

The spatial structure of the unstable eigenfunctions is shown in Fig.~\ref{fig1}(c--f). In both geometries, the perturbation of the polymer stretch, $\textrm{Tr}\,\bm{\delta c}$, is concentrated in thin sheets on either side of the centreline that are inclined with respect to the flow direction and form a chevron-like pattern. This is the linear precursor of the narwhal structures observed in direct numerical simulations of purely elastic channel and pipe flows \citep{Morozov2022,Morozov2025preprint,Burns2026preprint}. The perturbation vorticity, by contrast, is not localised and fills the whole cross-section, similar to the eigenfunctions of the Oldroyd-B centre mode at finite $\Re$ \citep{Chaudhary2021,Khalid2021a}. 

\begin{figure}[!t]
\begin{center}
\includegraphics[scale=\figscale]{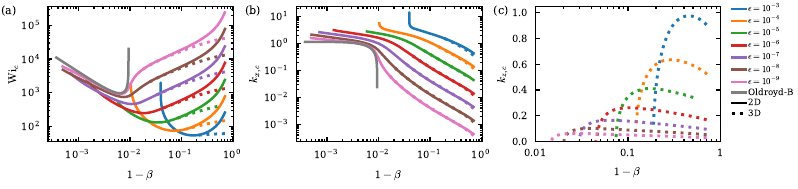}
\end{center}
\caption{Purely elastic ($\Re=0$) centre-mode instability in channel flow of an sPTT fluid. (a) Critical Weissenberg number, (b) critical streamwise wavenumber and (c) critical spanwise wavenumber as functions of $1-\beta$ for different values of $\epsilon$; the grey line is the Oldroyd-B fluid \citep{Khalid2021}. Solid lines: two-dimensional perturbations; dotted lines: three-dimensional perturbations, shown where they are the most unstable.}
\label{fig2}
\end{figure}

\subsection{Channel flow}\label{sec:channel}

Fig.~\ref{fig2} summarises the region of existence of the instability in purely elastic channel flow. For each value of $\epsilon$, we show the critical Weissenberg number, $\Wi_c$, minimised over the wavenumbers, as a function of $1-\beta$, together with the corresponding critical wavenumbers. In the limit $\epsilon\to 0$, our results recover the Oldroyd-B centre mode of \citet{Khalid2021}: an unstable region confined to $\beta>0.9905$, with a minimum $\Wi_c\approx 974$ at $\beta\approx 0.994$, and bounded on the side of lower $\beta$ by a long-wave cut-off where $k_{x,c}\to 0$ and $\Wi_c$ diverges.

Shear thinning changes this picture in two ways. Firstly, the long-wave cut-off at $\beta=0.9905$ disappears for any $\epsilon>0$ that we considered. Even for $\epsilon=10^{-9}$, where the critical curve is nearly indistinguishable from the Oldroyd-B one in the ultra-dilute regime, the instability continues smoothly past $\beta=0.9905$ to at least $\beta\approx 0.3$, albeit at rapidly increasing $\Wi_c$ and decreasing $k_{x,c}$. Secondly, as $\epsilon$ increases, the minimum of $\Wi_c$ moves to lower values of $\beta$ and Weissenberg numbers.
For the largest values of $\epsilon$, the most dilute solutions ($\beta\to1$) become stable, and for $\epsilon=10^{-4}$ and $10^{-3}$ we find no instability above $\beta\approx 0.99$ and $\beta\approx 0.96$, respectively. This boundary corresponds to a short-wave cut-off, where $k_{x,c}$ grows rapidly (Fig.~\ref{fig2}b). Shear thinning, therefore, does not simply lower the threshold of the Oldroyd-B instability, but moves the region of instability away from the ultra-dilute regime towards concentrations that are routinely used in experiments \citep{Pan2013,Samanta2013,Choueirie2021,Shnapp2022,Kamil2026preprint}.

Away from the ultra-dilute limit, the critical streamwise wavenumber decreases with increasing polymer concentration and, at fixed $\beta$, with decreasing $\epsilon$. In the latter case, $k_{x,c}$ and $\Wi_c$ change in such a way that their product remains approximately constant, indicating that the wavelength of the critical mode scales with the distance, $\lambda U$, travelled by the fluid over one relaxation time.

For sufficiently concentrated solutions, the most unstable perturbations are three-dimensional (dotted lines in Fig.~\ref{fig2}a,c). The critical spanwise wavenumber departs from zero in a close vicinity of $\beta\to1$, and remains finite at smaller $\beta$. Since an analogue of Squire's theorem guarantees that two-dimensional perturbations are the most unstable in an Oldroyd-B fluid \citep{Bistagnino_2007}, the preference for three-dimensional perturbations is a direct consequence of shear thinning \citep{Furlan2022}. The reduction of $\Wi_c$ associated with three-dimensionality is significant, and we observe large modifications to the region of existence of the centre-mode instability.

\begin{figure}[!t]
\begin{center}
\includegraphics[scale=\figscale]{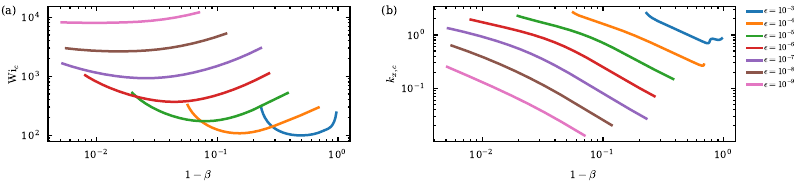}
\end{center}
\caption{Purely elastic ($\Re=0$) centre-mode instability in pipe flow of an sPTT fluid, axisymmetric perturbations ($m=0$). (a) Critical Weissenberg number and (b) critical axial wavenumber as functions of $1-\beta$ for different values of $\epsilon$. No instability exists for an Oldroyd-B fluid.}
\label{fig3}
\end{figure}

\subsection{Pipe flow}\label{sec:pipe}

The corresponding results for pipe flow are shown in Fig.~\ref{fig3}. In contrast to the channel, pipe flow of an Oldroyd-B fluid exhibits no centre-mode instability at $\Re=0$, and the existence of the purely elastic instability reported here relies entirely on shear thinning. For each value of $\epsilon$, the instability is found in a finite interval of $\beta$, with $\Wi_c$ attaining a minimum inside it. As $\epsilon$ decreases, this interval and the location of the minimum move towards dilute solutions, while $\Wi_c$ increases. For the strongest shear thinning considered, $\epsilon=10^{-3}$, the instability is present from $\beta\approx 0.77$ down to the most concentrated solutions that we considered. As in channel flow, the critical wavenumber decreases with increasing concentration and with decreasing $\epsilon$, with $k_{x,c}\Wi_c$ approximately independent of $\epsilon$ at fixed $\beta$. We have found no unstable non-axisymmetric modes.

\begin{figure}[!t]
\begin{center}
\includegraphics[scale=1.6]{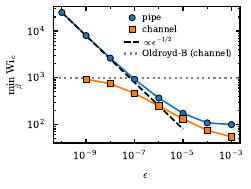}
\end{center}
\caption{Lowest critical Weissenberg number over all values of $\beta$ as a function of $\epsilon$ for $\Re=0$: (circles) pipe flow, where the dashed line shows $\Wi\propto\epsilon^{-1/2}$; (squares) channel flow, where the dotted line is the Oldroyd-B value, $\Wi_c=974$ \citep{Khalid2021}.}
\label{fig4}
\end{figure}

\subsection{The Oldroyd-B limit}\label{sec:OBlimit}

The different roles played by shear thinning in the two geometries are most clearly seen in Fig.~\ref{fig4}, where we show the lowest critical Weissenberg number over all values of $\beta$ as a function of $\epsilon$. In channel flow, $\min_\beta \Wi_c$ decreases monotonically with $\epsilon$ and, as $\epsilon\to 0$, tends to its Oldroyd-B value \citep{Khalid2021}. Shear thinning here only facilitates an instability that already exists in its absence. In pipe flow, on the other hand, $\min_\beta\Wi_c$ diverges as $\epsilon^{-1/2}$ for $\epsilon\lesssim 10^{-7}$,
consistent with the absence of the instability in the Oldroyd-B limit. Since the laminar velocity profile and the shear stress depend on $\epsilon$ and $\Wi$ only through the combination $\Wi\sqrt{\epsilon}$, this scaling suggests that the purely elastic instability in pipe flow sets in at a fixed level of shear thinning of the base flow, $\Wi_c\sqrt{\epsilon}\approx 0.25$. At these conditions, the polymer contribution to the viscosity at the wall is reduced by about a quarter of its zero-shear value. Shear thinning that only moderately affects the laminar flow is, therefore, sufficient to render viscoelastic pipe flow linearly unstable in the absence of inertia, while in the Oldroyd-B limit the threshold is pushed to infinity.

\subsection{Connection to the elasto-inertial centre mode}\label{sec:Re}

\begin{figure}[!t]
\begin{center}
\includegraphics[scale=\figscale]{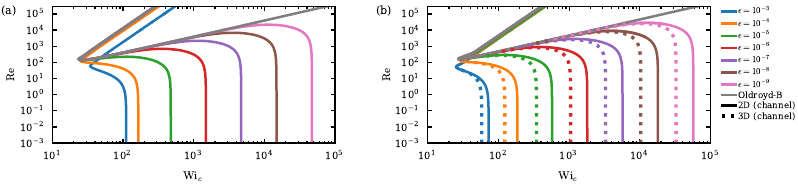}
\end{center}
\caption{Neutral stability curves in the $(\Wi,\Re)$ plane at $\beta=0.65$ for (a) pipe flow, $m=0$, and (b) channel flow, for different values of $\epsilon$ and for the Oldroyd-B fluid. In (b), solid and dotted lines correspond to two- and three-dimensional perturbations, respectively. The flow is unstable to the right of the vertical lines of constant $\Wi$, at low $\Re$, and inside the wedge-shaped region, at $\Re\gtrsim 10^2$.}
\label{fig5}
\end{figure}

Finally, we examine how the purely elastic instability connects to the elasto-inertial centre mode at finite Reynolds numbers. Fig.~\ref{fig5} shows neutral curves in the $(\Wi,\Re)$ plane for $\beta=0.65$, a viscosity ratio at which the Oldroyd-B fluid is unstable only at $\Re\gtrsim 10^2$ in both geometries \citep{Chaudhary2021,Khalid2021a}. For the Oldroyd-B fluid, the unstable region is a wedge. On its low-$\Wi$ side, it is bounded
by a branch on which $\Wi_c \sim \Re^{1/3}$ \citep{Chaudhary2021,Kerswell2024}, and on its 
high-$\Wi$ side by a branch on which the elasticity number, $\Wi/\Re$, reaches its maximum value, $\Wi/\Re \approx 0.25$ in pipe flow. For larger $\Wi/\Re$ the Oldroyd-B fluid is stable.

For the sPTT fluid, at low $\Re$ the neutral curves comprise a branch of constant $\Wi_c$, independent of $\Re$, and the instability is purely elastic. The plateau value scales as $\epsilon^{-1/2}$, with $\Wi_c\sqrt{\epsilon}\approx 1.5$ in pipe flow and $\Wi_c\sqrt{\epsilon}\approx 1.81$ and $1.0$ for two- and three-dimensional perturbations in channel flow, respectively. At $\Re=O(1)$ to $O(10)$ and sufficiently large $\epsilon$, inertia starts to destabilise the flow, and the neutral curves bend towards lower $\Wi$, approaching the nose of the Oldroyd-B wedge at $\Re=O(10^2)$. 
For $\epsilon\lesssim 10^{-5}$, they then follow both branches of the Oldroyd-B neutral curve, except that the high-$\Wi$ branch no longer extends to arbitrarily large $\Re$: at a value of $\Re$ that grows as $\epsilon$ decreases, it turns back and joins the purely elastic plateau. For the largest $\epsilon=10^{-3}$, the minimum of the neutral curve moves to lower $\Re$, and its low-$\Wi$ branch lies at larger $\Wi$ than the Oldroyd-B one.

The purely elastic and the elasto-inertial instabilities therefore stem from the same origin and a single neutral surface connects the centre mode at $\Re=0$ to the centre mode of \citet{Garg2018} and \citet{Khalid2021a} at $\Re=O(10^2)$ to $O(10^3)$. The role of shear thinning is to remove the high-elasticity cut-off of the Oldroyd-B instability, so that flows that are stable for $\Wi/\Re$ above its maximum value become unstable again once $\Wi$ is large enough for the base flow to thin. This extends the continuous pathway found by \citet{Khalid2021} for ultra-dilute Oldroyd-B solutions in a channel to moderate $\beta$ and, separately, to pipe flow.

\section{Discussion}\label{sec:discussion}

Our results demonstrate that purely elastic channel and pipe flows of an sPTT fluid exhibit a centre-mode instability over a wide range of material parameters. While in channel flow of an Oldroyd-B fluid the instability is restricted to ultra-dilute solutions, $\beta\to1$, and very large values of the Weissenberg number, $\Wi\sim O(10^3)$ \citep{Khalid2021}, and is absent altogether in pipe flow \citep{Chaudhary2021}, in an sPTT fluid with $\epsilon\sim10^{-4}$--$10^{-3}$,  we find a centre-mode instability for $\beta\approx 0.5$--$0.9$ and $\Wi_c \sim O(10^2)$ in both geometries. Rather than a niche feature of the Oldroyd-B model that is likely beyond experimental reach, the centre-mode instability thus becomes accessible to modern experiments, provided the rheology of the fluid is accurately described by the sPTT model.

While it would be tempting to associate this behaviour with the shear-thinning properties of the sPTT model, absent in the Oldroyd-B constitutive equation, the situation is more subtle. In steady and homogeneous flows, the sPTT equations of motion can be mapped exactly onto the FENE-P model, upon the identification $\epsilon \leftrightarrow 1/L^2$ and $\Wi_\mathrm{sPTT} \leftrightarrow (1-3/L^2) \Wi_\mathrm{FENE-P}$, where $L$ is the finite-extensibility  parameter of the FENE-P model \citep{John2024}. The fact that the sPTT and FENE-P models give drastically different predictions regarding the existence of the centre-mode instability in parallel shear flows is thus likely to be rooted in the different transient response exhibited by these models, such as strong stress overshoots, and hence sharp stress gradients, inherent to FENE-P flows. Further work is needed to distil the  rheological properties that support the centre-mode instability. We note here that if our hypothesis regarding the stress overshoot is correct, we expect the Giesekus model to exhibit similar instabilities to the sPTT model studied here. 

Our observations raise a broader question of whether the phenomenology of elastic turbulence, its transition pathways, and its structures are different for different models. Recent work by   \citet{John2024b} showed that the FENE-P and FENE-CR models, on the one hand, and the sPTT and Giesekus models, on the other hand, give significantly different predictions for flows in complex geometries, which can be verified experimentally. The existing literature already offers some support for this idea: while some experiments showed a rather smooth transition from laminar to elastic \citep{Kamil2026preprint} and elasto-inertial turbulence \citep{Samanta2013,Choueirie2021}, others exhibited a sharp, sub-critical transition \citep{Pan2013,Shnapp2022}, although it might be difficult to distinguish between a true linear instability and a weakly sub-critical one. It is therefore likely that elastic turbulence exhibited by various fluids depends qualitatively on their transient rheological response, in experiments as well as in simulations, and a detailed classification of such differences is called for.

\bigskip
% Backsection
\noindent\textbf{Acknowledgements.}
We gratefully acknowledge discussions with Bj\"{o}rn Hof and Jacob Page. For the purpose of open access, the authors have applied a Creative Commons Attribution (CC BY) licence to any Author Accepted Manuscript version arising from this submission.

\noindent\textbf{Funding.}
This work was supported by a grant from the Simons Foundation International (SSRFA 6826, K.J.B.).

\noindent\textbf{Declaration of Interests.}
The authors report no conflict of interest.

\noindent\textbf{Use of artificial intelligence tools.}
Claude code (Anthropic; model Opus 5.5, accessed between 8 October and 9 November 2026) was used to optimise our existing numerical code for computational efficiency. This code was verified against the author's original code.

% Bibliography
\printbibliography

@article{Vasil2016,
  title = {Tensor calculus in polar coordinates using {J}acobi polynomials},
  volume = {325},
  journal = {J. Comput. Phys.},
  author = {Vasil, G. M. and Burns, K. J. and Lecoanet, D. and Olver, S. and Brown, B. P. and Oishi, J. S.},
  year = {2016},
  pages = {53–73}
}

@article{Lellep2024,
  title={Purely elastic turbulence in pressure-driven channel flows},
  author={Lellep, M. and Linkmann, M. and Morozov, A.},
  journal={Proc. Natl. Acad. Sci. U.S.A.},
  volume={121},
  number={9},
  pages={e2318851121},
  year={2024}
}

@article{Datta2022,
  title = {Perspectives on viscoelastic flow instabilities and elastic turbulence},
  author = {Datta, S. S. and Ardekani, A. M. and Arratia, P. E. and Beris, A. N. and Bischofberger, I. and McKinley, G. H. and Eggers, J. G. and L\'opez-Aguilar, J. E. and Fielding, S. M. and Frishman, A. and Graham, M. D. and Guasto, J. S. and Haward, S. J. and Shen, A. Q. and Hormozi, S. and Morozov, A. and Poole, R. J. and Shankar, V. and Shaqfeh, E. S. G. and Stark, H. and Steinberg, V. and Subramanian, G. and Stone, H. A.},
  journal = {Phys. Rev. Fluids},
  volume = {7},
  pages = {080701},
  year = {2022}
}

@article{Burns2020,
author = {{Burns}, K. J. and {Vasil}, G. M. and {Oishi}, J. S. and {Lecoanet}, D. and {Brown}, B. P.},
title = "{Dedalus: A flexible framework for numerical simulations with spectral methods}",
journal = {Phys. Rev. Res.},
year = {2020},
volume = {2},
pages = {023068}
}

@article{Page2020,
  title = {Exact Traveling Wave Solutions in Viscoelastic Channel Flow},
  author = {Page, J. and Dubief, Y. and Kerswell, R. R.},
  journal = {Phys. Rev. Lett.},
  volume = {125},
  issue = {15},
  pages = {154501},
  year = {2020}
}

@article{Pan2013,
Author = {Pan, L. and Morozov, A. and Wagner, C. and Arratia, P. E.},
Title = {Nonlinear Elastic Instability in Channel Flows at Low {R}eynolds Numbers},
Journal = {Phys. Rev. Lett.},
Year = {2013},
Volume = {110},
pages = {174502}
}

@Article{Buza2022,
  author    = {Buza, G. and Page, J. and Kerswell, R. R.},
  title     = {Weakly nonlinear analysis of the viscoelastic instability in channel flow for finite and vanishing {R}eynolds numbers},
  journal   = {J. Fluid Mech.},
  year      = {2022},
  volume    = {940},
  pages     = {A11}
}

@article{Buza2022a, 
author={Buza, G. and Beneitez, M. and Page, J. and Kerswell, R. R.}, 
title={Finite-amplitude elastic waves in viscoelastic channel flow from large to zero {R}eynolds number}, 
journal={J. Fluid Mech.}, 
volume={951}, 
year={2022}, 
pages={A3}
}

@article{Beneitez2024,
  title = {Multistability of elasto-inertial two-dimensional channel flow},
  author = {Beneitez, M. and Page, J. and Dubief, Y. and Kerswell, R. R.},
  journal={J. Fluid Mech.}, 
  volume={981}, 
  year={2024}, 
  pages={A30}
}

@article{Kerswell2024, 
author={Kerswell, R. R. and Page, J.}, 
title={Asymptotics of the centre-mode instability in viscoelastic channel flow: with and without inertia}, 
journal={J. Fluid Mech.}, 
volume={991},
pages={A13},
year={2024} 
}

@article{Lewy2025,
author={Lewy, T. and Kerswell, R. R.}, 
title={Revisiting two-dimensional viscoelastic {K}olmogorov flow: a centre-mode-driven transition},
journal={J. Fluid Mech.}, 
volume={1007},
pages={A55},
year={2025}, 
}

@article{Zhu2026, 
title={On the essential structure of exact travelling-wave solutions in viscoelastic flow}, volume={1033}, 
journal={J. Fluid Mech.}, 
author={Zhu, L. and Kerswell, R. R.}, 
year={2026}, 
pages={R6}
}

@article{Khalid2021,
  title = {Continuous Pathway between the Elasto-Inertial and Elastic Turbulent States in Viscoelastic Channel Flow},
  author = {Khalid, M. and Shankar, V. and Subramanian, G.},
  journal = {Phys. Rev. Lett.},
  volume = {127},
  pages = {134502},
  year = {2021}
}

@article{Khalid2021a, 
title={The centre-mode instability of viscoelastic plane {P}oiseuille flow}, 
volume={915}, 
journal={J. Fluid Mech.}, 
author={Khalid, M. and Chaudhary, I. and Garg, P. and Shankar, V. and Subramanian, G.}, 
year={2021}, 
pages={A43}
}

@article {Choueirie2021,
  author = {Choueiri, G. H. and Lopez, J. M. and Varshney, A. and Sankar, S. and Hof, B.},
  title = {Experimental observation of the origin and structure of elastoinertial turbulence},
  volume = {118},
  number = {45},
  year = {2021},
  journal = {Proc. Natl. Acad. Sci. U.S.A.}
}

@ARTICLE{Samanta2013,
  author = {Samanta, D. and Dubief, Y. and Holzner, M. and Schäfer, C. and Morozov, A. N. and Wagner, C. and Hof, B.},
  title = {Elasto-inertial turbulence},
  journal = {Proc. Natl. Acad. Sci. U.S.A.},
  year = {2013},
  volume = {110},
  pages = {10557–10562},
  number = {26}
}

@misc{Kamil2026preprint,
      title={Low inertia limit of elasto-inertial turbulence}, 
      author={Kamil, S. and Suresh, S. S. and Lopez, J. M. and Hof, B.},
      year={2026},
      eprint={2609.11581},
      archivePrefix={arXiv}
}

@article{PhanThien1977,
title = {A new constitutive equation derived from network theory},
author = {Phan-Thien, N. and Tanner, R. I.},
journal = {J. Non-Newton. Fluid Mech.},
volume = {2},
pages = {353-365},
year = {1977},
}

@article{Sanchez2022,
title = {Understanding viscoelastic flow instabilities: {O}ldroyd-{B} and beyond},
journal = {J. Non-Newton. Fluid Mech.},
volume = {302},
pages = {104742},
year = {2022},
issn = {0377-0257},
doi = {https://doi.org/10.1016/j.jnnfm.2022.104742},
url = {https://www.sciencedirect.com/science/article/pii/S0377025722000015},
author = {{Castillo S\'{a}nchez}, H. A. and Jovanovi\'{c}, M. R. and Kumar, S. and Morozov, A. and Shankar, V. and Subramanian, G. and Wilson, H. J.}
}

@ARTICLE{McKinley1996,
  author = {McKinley, G. H. and Pakdel, P. and Oeztekin, A.},
  title = {Rheological and geometric scaling of purely elastic flow instabilities},
  journal = {J. Non-Newton. Fluid Mech.},
  year = {1996},
  volume = {67},
  pages = {19--47}
}

@article{Shnapp2022,
  title = {Nonmodal elastic instability and elastic waves in weakly perturbed channel flow},
  author = {Shnapp, R. and Steinberg, V.},
  journal = {Phys. Rev. Fluids},
  volume = {7},
  pages = {063901},
  year = {2022}
}

@Book{Bird1987,
author = {Bird, R. B.},
title = {Dynamics of Polymeric Liquids: Fluid Mechanics },
publisher = {Wiley},
year = {1987},
volume = {1},
address = {New York},
edition = {2}
}

@ARTICLE{Shaqfeh1996,
  author = {Shaqfeh, E. S. G.},
  title = {Purely Elastic Instabilities in Viscometric Flows},
  journal = {Annu. Rev. Fluid Mech.},
  year = {1996},
  volume = {28},
  pages = {129--185},
  number = {1}
}

@article{Morozov2022,
  title = {Coherent Structures in Plane Channel Flow of Dilute Polymer Solutions with Vanishing Inertia},
  author = {Morozov, A.},
  journal = {Phys. Rev. Lett.},
  volume = {129},
  pages = {017801},
  year = {2022}
}

@article{Lellep2023, 
title={Linear stability analysis of purely elastic travelling-wave solutions in pressure-driven channel flows}, 
author={Lellep, M. and Linkmann, M. and Morozov, A.},
volume={959}, 
journal={J. Fluid Mech.}, 
year={2023}, 
pages={R1}
}

@article{Smits2011,
author = {Smits, A. J. and McKeon, B. J. and Marusic, I.},
title = {High–{R}eynolds Number Wall Turbulence},
journal = {Annu. Rev. Fluid Mech.},
volume = {43},
pages = {353-375},
year = {2011}
}

@Article{Wilson1999,
  author    = {Wilson, H. J. and Renardy, M. and Renardy, Y.},
  title     = {Structure of the spectrum in zero {R}eynolds number shear flow of the {UCM} and {O}ldroyd-{B} liquids},
  journal   = {J. Non-Newton. Fluid Mech.},
  year      = {1999},
  volume    = {80},
  pages     = {251--268}
}

@article{Berti2008,
  title = {Two-dimensional elastic turbulence},
  author = {Berti, S. and Bistagnino, A. and Boffetta, G. and Celani, A. and Musacchio, S.},
  journal = {Phys. Rev. E},
  volume = {77},
  pages = {055306},
  year = {2008}
}

@article{Berti2010,
  title = {Elastic waves and transition to elastic turbulence in a two-dimensional viscoelastic {K}olmogorov flow},
  author = {Berti, S. and Boffetta, G.},
  journal = {Phys. Rev. E},
  volume = {82},
  pages = {036314},
  year = {2010}
}

@article{Garg2018,
  title = {Viscoelastic Pipe Flow is Linearly Unstable},
  author = {Garg, P. and Chaudhary, I. and Khalid, M. and Shankar, V. and Subramanian, G.},
  journal = {Phys. Rev. Lett.},
  volume = {121},
  pages = {024502},
  year = {2018}
}

@article{Chaudhary2021, 
title={Linear instability of viscoelastic pipe flow}, 
volume={908}, 
journal={J. Fluid Mech.}, 
author={Chaudhary, I. and Garg, P. and Subramanian, G. and Shankar, V.}, 
year={2021}, 
pages={A11}}

@article{Rota2024,
  title = {Unified view of elastic and elasto-inertial turbulence in channel flows at low and moderate {R}eynolds numbers},
  author = {Foggi Rota, G. and Amor, C. and Le Clainche, S. and Rosti, M. E.},
  journal = {Phys. Rev. Fluids},
  volume = {9},
  pages = {L122602},
  year = {2024}
}

@article{Lin2025,
title = {{GPU} acceleration of a hi-fidelity algorithm for direct numerical simulation of polymer-induced/modified turbulence},
journal = {J. Non-Newton. Fluid Mech.},
volume = {342},
pages = {105437},
year = {2025},
author = {Lin, F. and Liao, Z.-M. and Zhao, Z. and Liu, N. and Lu, X.-Y. and Khomami, B.}
}

@article{Nichols2025,
  title = {Period-doubling route to chaos in viscoelastic {K}olmogorov flow},
  author = {Nichols, J. and Guy, R. D. and Thomases, B.},
  journal = {Phys. Rev. Fluids},
  volume = {10},
  pages = {L041301},
  year = {2025}
}

@article{Goffin2026,
  title = {Follow the curvature of viscoelastic stress: Insights into the steady arrowhead structure},
  author = {Goffin, P.-Y. and Dubief, Y. and Terrapon, V. E.},
  journal = {Phys. Rev. Fluids},
  volume = {11},
  pages = {L061301},
  year = {2026}
}

@misc{Burns2026preprint,
      title={Narwhals in pipes}, 
      author={Burns, K. J. and Oishi, J. S. and Skene, C. S. and Lecoanet, D. and Brown, B. P. and Linkmann, M. and Vasil, G. M. and Morozov, A.},
      year={2026},
      howpublished={(in preparation)}
}

@misc{Morozov2025preprint,
  title={Narwhals and their blessings: exact coherent structures of elastic turbulence in channel flows}, 
  author={Morozov, A. and Lellep, M. and Capocci, D. and Linkmann, M.},
  year={2025},
  eprint={2509.03175},
  archivePrefix={arXiv},
  primaryClass={physics.flu-dyn},
  url={https://arxiv.org/abs/2509.03175}, 
}

@article{Furlan2022,
title = {On the validity of {S}quire's theorem for viscoelastic fluid flows},
journal = {J. Non-Newton. Fluid Mech.},
volume = {307},
pages = {104880},
year = {2022},
author = {da Silva Furlan, L. J. and {de Mendonca}, M. T. and {de Araujo}, M. T. and {de Souza}, L. F.}
}

@article{Tej2026, 
title={Transition in elastic {D}ean flow: the centre-mode versus hoop-stress pathways}, 
volume={1043}, 
journal={J. Fluid Mech.}, 
author={Tej, P. S. D. S. P. and Subramanian, G. and Shankar, V.}, 
year={2026}, 
pages={A8}
}

@article{Khalid2025, 
title={Role of finite extensibility on the centre-mode instability in viscoelastic channel flow}, 
volume={1009}, 
journal={J. Fluid Mech.}, 
author={Khalid, M. and Badoni, A. and Dutta, D. and Naidu, P. and Subramanian, G. and Shankar, V.}, 
year={2025}, 
pages={A28}
}

@article{John2024, 
title={A comparison between the {FENE-P} and {sPTT} constitutive models in large-amplitude oscillatory shear}, 
volume={979}, 
journal={J. Fluid Mech.}, 
author={John, T. P. and Poole, R. J. and Kowalski, A. J. and Fonte, C. P.}, 
year={2024}, 
pages={A10}
}

@article{Oliveira1999, 
title={Analytical solution for fully developed channel and pipe flow of {P}han-{T}hien–{T}anner fluids}, 
volume={387}, 
journal={J. Fluid Mech.}, 
author={Oliveira, P. J. and Pinho, F. T.}, 
year={1999}, 
pages={271–280}
}

@article{Cruz2005,
title = {Analytical solutions for fully developed laminar flow of some viscoelastic liquids with a {N}ewtonian solvent contribution},
journal = {J. Non-Newton. Fluid Mech.},
volume = {132},
pages = {28-35},
year = {2005},
author = {Cruz, D. O. A. and Pinho, F. T. and Oliveira, P. J.}
}

@article{John2024b,
title = {Comparing flows of {FENE-P}, {sPTT}, and {G}iesekus model fluids in a helical static mixer},
journal = {J. Non-Newton. Fluid Mech.},
volume = {329},
pages = {105249},
year = {2024},
author = {John, T. P. and Stewart, J. T. and Poole, R. J. and Kowalski, A. and Fonte, C. P.}
}

@article{Bistagnino_2007, 
title={Nonlinear dynamics of the viscoelastic {K}olmogorov flow}, 
volume={590}, 
DOI={10.1017/S0022112007007859}, 
journal={J. Fluid Mech.}, 
author={Bistagnino, A. and Boffetta, G. and Celani, A. and Mazzino, A. and Puliafito A. and Vergassola, M.}, 
year={2007}, 
pages={61–80}
}

\end{document}